\documentclass[preprint,12pt,prd,aps,
amssymb]{revtex4}

\usepackage{feynmf}

\def\bfsigma{\mbox{\boldmath $\sigma$}}

\newcommand{\nn}{\nonumber}
\newcommand{\be}{\begin{equation}}
\newcommand{\ee}{\end{equation}}
\newcommand{\bea}{\begin{eqnarray}}
\newcommand{\eea}{\end{eqnarray}}

\def\siml{{\ \lower-1.2pt\vbox{\hbox{\rlap{$<$}\lower6pt\vbox{\hbox{$\sim$}}}}\ }}

\newcommand{\Appendix}[1]%
    {%
     \section{#1}%
      }

\begin{document}
\begin{fmffile}{metafont}

\preprint{\small \tt
 UB-ECM-PF 06/28
\vspace{2cm}}
\title{\bf The nucleon-nucleon potential beyond the static approximation}
\author {Jorge Mondejar, Joan
  Soto}
\affiliation{Deptartament d'Estructura i Constituents de la Mat\`eria,
  Universitat de Barcelona \\ Diagonal 647, 08028 Barcelona, Catalonia, Spain}
\begin{abstract}

\noindent
We point out that, due to the use of static nucleon propagators in Heavy Baryon Chiral Perturbation Theory, the current calculations of the nucleon-nucleon potential miss certain contributions starting at two loops. These contributions give rise to contact interactions, which are both parametrically and numerically more important than the so called NNLO potentials.
They show a peculiar dependence on the light quark masses, which should be taken into account when performing chiral extrapolations of lattice data. However, they do not appear to have an impact on phenomenology since they can be absorbed into redefinitions of unknown parameters which are usually fitted to data.

\end{abstract}

\maketitle

\section{Introduction}

Since Weinberg's pioneering work in the early 90's \cite{Weinberg:1990rz,Weinberg:1991um}, there has been an enormous development of effective theory methods for few nucleon systems (see \cite{Bedaque:2002mn,Epelbaum:2005pn,Hammer:2006qj} for recent reviews). However, there is still a lack of consensus on how calculations in the so called NN effective theory including pions must be organized \cite{Epelbaum:2006pt,Epelbaum:2004fk,Epelbaum:1999dj,Epelbaum:1998ka,Ordonez:1992xp,Ordonez:1993tn,Ordonez:1995rz}\cite{Kaplan:1998tg,Kaplan:1998we,Fleming:1999bs,Fleming:1999ee}\cite{Nogga:2005hy,Beane:2001bc}\cite{Valderrama:2005ku,PavonValderrama:2005uj,Valderrama:2005wv,PavonValderrama:2005gu,PavonValderrama:2004nb,PavonValderrama:2003np,Nieves:2003uu}\cite{Djukanovic:2006mc,Gegelia:2004pz,Gegelia:2001ev,Gegelia:1999gf,Gegelia:1998ee}\cite{Timoteo:2005ia,Frederico:1999ps}\cite{Yang:2006ix,Yang:2004mq,Yang:2004ss,Yang:2004zg}\cite{Birse:2005um}\cite{Oller:2003px}\cite{Eiras:2001hu}. Starting from the Heavy Baryon Chiral Lagrangian (HB$\chi$L)\cite{Jenkins:1990jv}, one is interested in building a low energy effective theory for nucleon energies much lower than the pion mass.
It seems clear that the remaining effective theory must consist of nucleons interacting through a potential. Hence the program may be divided in two: (i) calculating the potential and (ii) organizing the remaining quantum mechanical calculation. We shall not enter in point (ii) here, which is where the main difficulties arise. Concerning point (i), following the so called Weinberg approach, it is commonly believed that one can use HB$\chi$PT counting rules and the outcome of the calculation can be easily organized in standard chiral counting in $1/\Lambda_\chi$; $\Lambda_\chi\sim 4\pi F_\pi$, $F_\pi$ being the pion decay constant. We point out here that this is not so. If we understand the potentials as matching coefficients which arise after integrating out higher energy degrees of freedom \cite{Pineda:1997bj,Eiras:2001hu}, then, starting at two loops, there are contributions to them which are missed if the static approximation is used for the nucleon propagators, as prescribed by HB$\chi$PT counting rules
\footnote{
The static approximation also fails for certain observables in the single nucleon sector \cite{Becher:1999he}.}. 
This is due to the fact that the energy scale given by the pion mass $m_\pi$ has an associated three-momentum scale $\sqrt{m_\pi M_N}$, $M_N$ being the nucleon mass, which may make the kinetic term of the nucleon propagator as important as the energy when integrating out degrees of freedom of energy $m_\pi$
\footnote{An analogous situation has been identified in the contexts of hadronic atoms \cite{Eiras:1999xx} and heavy quarkonium \cite{Brambilla:2003mu}.}. This becomes apparent when individual Feynman diagrams are analyzed with the method of the strategy of regions or threshold expansions \cite{Beneke:1997zp,Smirnov:2002pj} with non-relativistic nucleon propagators, as will be shown in the next section. This kind of contributions are related to the so called radiation pions discussed in Refs. \cite{Fleming:1999bs,Fleming:1999ee}  within the KSW approach\cite{Kaplan:1998tg,Kaplan:1998we}. They are also related to the fractional powers of the pion mass which appear in pion production off nucleon-nucleon systems \cite{Cohen:1995cc,Hanhart:2000gp,Hanhart:2002bu,Baru:2004kw,Lensky:2005hb,Lensky:2005jc}. However, to our knowledge, they have never been discussed in the framework of the original Weinberg proposal for calculating the potential\cite{Weinberg:1990rz,Weinberg:1991um}. We present in section \ref{secdr} their complete contribution at two loops. In order to obtain it, the leading order lagrangian (in the isopin limit) augmented by the kinetic terms of the nucleon suffices, 
\be
\mathcal{L}=\mathcal{L}_{\pi}+\mathcal{L}_{\pi N}+\mathcal{L}_{N N}
\ee
The purely pionic sector reads
\begin{equation}
\mathcal{L}_{\pi}=\frac{F_{\pi}^2}{4}\left\{Tr[\nabla_{\mu}\rm{U}^{\dagger}\nabla^{\mu}\rm{U}+
m_\pi^2(\rm{U}+\rm{U}^{\dagger})]\right\} 
\ , \ U=e^{i\frac{\pi^a\tau^a}{F_{\pi}}} \ ,
\end{equation}
$\pi^a$ is the pion field and $\tau^a$ the (isospin) Pauli matrices. The pion-nucleon sector reads
\begin{equation}
\mathcal{L}_{\pi N}=
N^\dagger\left(iD_0-g_A({\bf u} \cdot \frac{\bfsigma}{2})+\frac{{\bf D}^2}{2M_N}\right)
N\ . \label{l2}
\end{equation}
$N$ is the nucleon field, $g_A$ is the axial vector coupling constant of the nucleon, 
$\bfsigma$ the (spin) Pauli matrices, and, defining $u^2(x)=U(x)$, the covariant derivative of the nucleon field is given by
\begin{equation}
D_{\mu}N=\left(\partial_{\mu}+\frac{1}{2}[u^{\dagger},\partial_{\mu}u]\right)N \ ,
\end{equation}
and the axial-vector type object $u_{\mu}$ 
by,
\begin{equation}
u_{\mu}=i(u^{\dagger}\nabla_{\mu}u-u\nabla_{\mu}u^  {\dagger})=i\{u^{\dagger},\nabla_{\mu}u\}=iu^{\dagger}\nabla_{\mu}Uu^{\dagger} \ .
\end{equation}
Finally, the nucleon-nucleon sector reads
\be
\mathcal{L}_{N N}=-{C_S\over 2}N^\dagger N N^\dagger N - {C_T\over 2} N^\dagger\bfsigma N N^\dagger\bfsigma N
\label{NN}
\ee
where $C_S$ and $C_T$ are low energy constants to be determined from the experiment or from QCD, for instance by lattice simulations \cite{Fukugita:1994ve,Beane:2006mx}.
In section \ref{secd} we discuss the relevance of these contributions in higher order calculations and the significance of our results.
Section \ref{secc} is devoted to the conclusions. Details of our calculation are presented in the Appendix.

\section{Threshold expansions: a sample calculation}
\label{secth}

In this section we will introduce the method of calculation that we
will apply. Many processes in physics, like those involving heavy
quarks or our pion-nucleon interaction, involve more than one mass
scale. Such processes are notoriously difficult to calculate in
perturbation theory beyond the one-loop level. To proceed one has to
resort to approximations, either numerical or analytical. Among the
latter finds its place the strategy of regions \cite{Beneke:1997zp,Smirnov:2002pj}. The idea is to
perform an asymptotic expansion of the integrals in certain ratios
of mass scales, so that the resulting integrals appearing in the
calculation of every term in the expansion are simpler, and the
expansion is homogeneous (each integral appearing in the
construction contributes to a determinate power in the expansion
parameter). In short, the method goes as follows:
\begin{enumerate}
\item{Determine the large and small scales in the problem.}
\item{Introduce factorization scales $\mu_i$ and divide the loop integration domain into regions in which the loop momentum is of the order of one of the scales in the problem.}
\item{Perform, in every region, a Taylor expansion in the parameters which are small in the given region, and stay at leading order. At this point, keep only the relevant regions and discard the rest: the relevant regions are those that somehow maintain the structure of poles of the original integral. If we integrate over multiple momenta, and the integrand has several propagators, so that we have one or more poles associated with each momentum, after performing the Taylor expansion we should still have at least one pole for each momentum. If we end up loosing all the poles that were associated with one of the variables of integration, the region we are considering is irrelevant and must not be taken into account. }
\item{After expansion, ignore all factorization scales and integrate over the entire loop integration domain in every relevant region.}
\end{enumerate}
The non-trivial point to justify is 4, which also guarantees the
homogeneity of the expansion formula. In order for that point to
hold it is essential to use dimensional (or analytic) regularization
for the integral, even if it is finite in four dimensions. Loosely
speaking, 4 follows in dimensional regularization from the property
that all integrals without scale vanish, but the truth is that at
present day there are no mathematical proofs of the method of
regions. The best we can say is that it has not failed yet, giving
asymptotic expansions for any diagram in any limit and having been
checked in numerous examples when comparing results of expansion
with existing explicit analytical results.
\bigskip

We will now show how this method works and how it enables us to
find new contributions to the nucleon-nucleon potential.
Consider the following diagram,
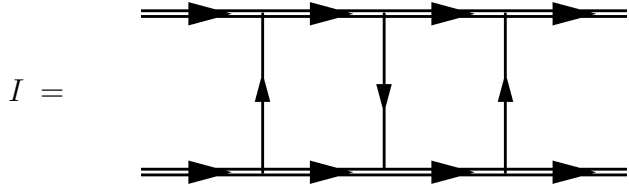
\begin{eqnarray}
\nonumber \\
\nonumber \\
I&=&
\parbox{30mm}{
\begin{fmfgraph*}(230,60)
\fmfleft{i1,i2} \fmfright{o1,o2}
\fmf{heavy,label=$p_1$,label.side=left}{i2,v1}
\fmf{heavy,label=$p_1+k$,label.side=left}{v1,v2}
\fmf{heavy,label=$p_1+k-l$,label.side=left}{v2,v3}
\fmf{heavy,label=$p_3$,label.side=left}{v3,o2}
\fmf{heavy,label=$p_2$,label.side=right}{i1,u1} 
\fmf{heavy,label=$p_2-k$,label.side=right}{u1,u2}
\fmf{heavy,label=$p_2-k+l$,label.side=right}{u2,u3} 
\fmf{heavy,label=$p_4$,label.side=right}{u3,o1}
\fmf{fermion,label=$k$,label.side=left,tension=0}{u1,v1}
\fmf{fermion,label=$l$,label.side=left,tension=0}{v2,u2}
\fmf{fermion,label=$p_2-p_4-k+l$,label.side=right,tension=0}{u3,v3}
\end{fmfgraph*}}
\nonumber \\
\nonumber \\
\nonumber \\
\nonumber \\
&=&i\left(\frac{g_A}{F_{\pi}}\right)^6(7\vec{\tau}_1\cdot\vec{\tau}_2-6)\frac{\sigma_1^i}{2}\frac{\sigma_1^j}{2}\frac{\sigma_1^l}{2}\frac{\sigma_2^r}{2}\frac{\sigma_2^s}{2}\frac{\sigma_2^t}{2}\int\frac{d^dk}{(2\pi)^d}\int\frac{d^dl}{(2\pi)^d} k^l k^t l^j l^s (k-l+p-p')^i (k-l+p-p')^r  \nonumber \\
& &\times\frac{1}{(k-l+\vec{p}-\vec{p}')^2-m_{\pi}^2+i\epsilon}\quad\frac{1}{l^2-m_{\pi}^2+i\epsilon}\quad\frac{1}{k^2-m_{\pi}^2+i\epsilon}\quad\frac{1}{k^0+p^0-\frac{(\vec{k}+\vec{p})^2}{2M_N}+i\epsilon} \nonumber \\
&
&\times\frac{1}{k^0-l^0+p^0-\frac{(\vec{k}-\vec{l}+\vec{p})^2}{2M_N}+i\epsilon}\quad\frac{1}{k^0-p^0+\frac{(\vec{k}+\vec{p})^2}{2M_N}-i\epsilon}\quad\frac{1}{k^0-l^0-p^0+\frac{(\vec{k}-\vec{l}+\vec{p})^2}{2M_N}-i\epsilon}
\ , \nonumber
\end{eqnarray}
\begin{equation}
\end{equation}
where we have put ourselves in the center of mass frame, that is,
\begin{equation}
\begin{array}{l}
p_1=(p^0,\vec{p}) \\
p_2=(p^0,-\vec{p})
\end{array}
\ , \quad
\begin{array}{l}
p_3=(p^0,\vec{p}') \\
p_4=(p^0,-\vec{p}')
\end{array} \ ,
\end{equation}
and so we write $p_1-p_3 = (0,\vec{p}-\vec{p}')$ as just $\vec{p}-\vec{p}'$. We will use this notation whenever the time component of a four-vector is zero or neglected.
\bigskip

We will assume that the nucleon momentum ${\bf p}$ and ${\bf p'}$, and the momentum transfer ${\bf p-p'}$ are of order $m_\pi$ whereas the nucleon energy $p_0$ is of order $ m_\pi^2/M_N$, which fixes the scales of the effective theory. The relevant energy scales in the diagram are $ m_\pi^2/M_N$ and $m_\pi$ and the associated three-momentum scales $m_\pi$ and $\sqrt{m_\pi M_N}$ respectively. We will have in mind the philosophy described in \cite{Eiras:2001hu}: modes with energies larger than $ m_\pi^2/M_N$ are integrated out giving rise to the potential. The strategy of regions will help us to separate these modes from the ones which must be kept in the effective theory (see \cite{Kniehl:2002br} for an example of such a calculation). 
Now we must break the integration domain into pieces. The relevant
regions are easily found by inspecting the poles of the propagators:
they are the parts of the integration domain in which one or more of
the propagators inside the integral develop a pole. Let us display next the leading term
of the Taylor expansion of the integrand in several regions:
\begin{itemize}
\item{
\begin{equation}
\left\{
\begin{array}{lll}
k^0\sim m_{\pi} \  , & \vec{k}\sim m_{\pi}\\
& & k^0-l^0\sim m_{\pi} \\
l^0\sim m_{\pi} \  , & \vec{l}\sim m_{\pi}\\
\end{array}
\right.
\end{equation}

\begin{equation}
\frac{(k-l+p-p')^i(k-l+p-p')^r}{(k-l+\vec{p}-\vec{p}')^2-m_{\pi}^2+i\epsilon}\quad\frac{l^jl^s}{l^2-m_{\pi}^2+i\epsilon}\quad\frac{k^lk^t}{k^2-m_{\pi}^2+i\epsilon}
\end{equation}
\begin{displaymath}
\times\frac{1}{k^0+i\epsilon}\quad\frac{1}{k^0-l^0+i\epsilon}\quad\frac{1}{k^0-i\epsilon}\frac{1}{k^0-l^0-i\epsilon} 
\ .
\end{displaymath}
This is a pure three pion exchange potential contribution, the same
we find in the static approximation. Strictly speaking, one should make sense of the pinch singularities above. There are several ways to treat them in the literature (see for instance \cite{Pineda:1998kn,Kniehl:2002br,Manohar:2006nz}). In the EFT framework we are working in they have a natural interpretation, since the same singularities appear in the lower energy EFT, which arise from the iterations of lower order potentials in the static limit \cite{Pineda:1998kn,Brambilla:2004jw}. Once these iterations are subtracted, a well defined expression is obtained. In this particular case, the needed subtractions are the iterations of an OPE (one pion exchange)- TPE (two pion exchange), TPE-OPE and OPE-OPE-OPE. The same comment holds for the various pinch singularities we will encounter below.}
\item{
\begin{equation}
\left\{
\begin{array}{lll}
k^0\sim m_{\pi} \  , & \vec{k}\sim m_{\pi}\\
& & k^0-l^0\sim \frac{m_{\pi}^2}{M_N}  \\
l^0\sim m_{\pi} \  , & \vec{l}\sim m_{\pi}\\
\end{array}
\right.
\end{equation}

\begin{equation}
\frac{(k-l+p-p')^i(k-l+p-p')^r}{-(\vec{k}-\vec{l}+\vec{p}-\vec{p}')^2-m_{\pi}^2}\quad\frac{l^jl^s}{l^2-m_{\pi}^2+i\epsilon}\quad\frac{k^lk^t}{k^2-m_{\pi}^2+i\epsilon}
\end{equation}
\begin{displaymath}
\times\frac{1}{k^0+i\epsilon}\quad\frac{1}{k^0-i\epsilon}\quad\frac{1}{k^0-l^0+p^0-\frac{(\vec{k}-\vec{l}+\vec{p})^2}{2M_N}+i\epsilon}\quad\frac{1}{k^0-l^0-p^0+\frac{(\vec{k}-\vec{l}+\vec{p})^2}{2M_N}-i\epsilon}
\ .
\end{displaymath}
We recognize here the contribution of a two pion exchange potential
followed by a one pion exchange potential, a contribution that also exists in the effective theory (i.e. in the quantum mechanical calculation) and hence, to
avoid double counting, must be discarded.}
\item{
\begin{equation}
\left\{
\begin{array}{llll}
k^0\sim \frac{m_{\pi}^2}{M_N} \  , & \vec{k}\sim m_{\pi}\\
& & & \\
l^0\sim m_{\pi} \  , & \vec{l}\sim m_{\pi}\\
\end{array}
\right.
\end{equation}

\begin{equation}
\frac{(k-l+p-p')^i(k-l+p-p')^r}{(\vec{k}-l+\vec{p}-\vec{p}')^2-m_{\pi}^2+i\epsilon}\quad\frac{l^jl^s}{l^2-m_{\pi}^2+i\epsilon}\quad\frac{k^lk^t}{-\vec{k}^2-m_{\pi}^2}
\end{equation}
\begin{displaymath}
\times\frac{1}{k^0+p^0-\frac{(\vec{k}+\vec{p})^2}{2M_N}+i\epsilon}\quad\frac{1}{k^0-p^0+\frac{(\vec{k}+\vec{p})^2}{2M_N}-i\epsilon}\quad\frac{1}{l^0-i\epsilon}\quad\frac{1}{l^0+i\epsilon}
\ .
\end{displaymath}
As opposed to the former region, this one gives a one pion potential
followed by a two pion potential, and must be discarded likewise.}
\item{
\begin{equation}
\left\{
\begin{array}{llll}
k^0\sim \frac{m_{\pi}^2}{M_N} \  , & \vec{k}\sim m_{\pi}\\
& &  & \\
l^0\sim \frac{m_{\pi}^2}{M_N} \  , & \vec{l}\sim m_{\pi}\\
\end{array}
\right.
\end{equation}

\begin{displaymath}
\frac{(k-l+p-p')^i(k-l+p-p')^r}{-(\vec{k}-\vec{l}+\vec{p}-\vec{p}')^2-m_{\pi}^2}\quad\frac{l^jl^s}{-\vec{l}^2-m_{\pi}^2}\quad\frac{k^lk^t}{-\vec{k}^2-m_{\pi}^2}\quad\frac{1}{k^0+p^0-\frac{(\vec{k}+\vec{p})^2}{2M_N}+i\epsilon}
\end{displaymath}
\begin{equation}
\times\frac{1}{k^0-p^0+\frac{(\vec{k}+\vec{p})^2}{2M_N}-i\epsilon}\quad\frac{1}{k^0-l^0+p^0-\frac{(\vec{k}-\vec{l}+\vec{p})^2}{2M_N}+i\epsilon}\quad\frac{1}{k^0-l^0-p^0+\frac{(\vec{k}-\vec{l}+\vec{p})^2}{2M_N}-i\epsilon}
\ .
\end{equation}
This is a one pion potential iterated three times, which also exists in the effective theory, and hence is to be
dropped like the other contributions.}
\item{
\begin{equation}
\left\{
\begin{array}{llll}
k^0\sim m_{\pi} \  , & \vec{k}\sim \sqrt{M_Nm_{\pi}}\\
& &  k^0-l^0\sim m_{\pi} \ , & \vec{k}-\vec{l}\sim \sqrt{M_Nm_{\pi}}\\
l^0\sim m_{\pi} \  , & \vec{l}\sim \sqrt{M_Nm_{\pi}}\\
\end{array}
\right.
\end{equation}

\begin{equation}
\frac{(k-l)^i(k-l)^r}{-(\vec{k}-\vec{l})^2}\quad\frac{l^jl^s}{-\vec{l}^2}\quad\frac{k^lk^t}{-\vec{k}^2}\quad\frac{1}{k^0-\frac{\vec{k}^2}{2M_N}+i\epsilon}\quad\frac{1}{k^0+\frac{\vec{k}^2}{2M_N}-i\epsilon}
\end{equation}
\begin{displaymath}
\times\frac{1}{k^0-l^0-\frac{(\vec{k}-\vec{l})^2}{2M_N}+i\epsilon}\quad\frac{1}{k^0-l^0+\frac{(\vec{k}-\vec{l})^2}{2M_N}-i\epsilon}
\ .
\end{displaymath}
After performing the integrations over $k^0$ and $l^0$ we are left
with
\begin{equation}
\frac{(k-l)^i(k-l)^r}{-(\vec{k}-\vec{l})^2}\quad\frac{l^jl^s}{-\vec{l}^2}\quad\frac{k^lk^t}{-\vec{k}^2}\quad
M_N^2 \frac{1}{\vec{k}^2(\vec{k}-\vec{l})^2} \
.
\end{equation}
The remaining integrand has no scales, and
therefore vanishes when we integrate over the remaining $d-1$
dimensions for $k$ and $l$ using dimensional regularization. 
}
\end{itemize}
By analogous arguments, we find that 
all the remaining 
regions 
have also vanishing contributions, except for the following one.

\begin{itemize}
\item{
\begin{equation}
\left\{
\begin{array}{llll}
k^0\sim m_{\pi} \  , & \vec{k}\sim \sqrt{M_Nm_{\pi}}\\
& &  k^0-l^0\sim m_{\pi}  & \\
l^0\sim m_{\pi} \  , & \vec{l}\sim m_{\pi}\\
\end{array}
\right.
\end{equation}

\begin{equation}
\frac{k^ik^t}{-\vec{k}^2}\quad\frac{l^jl^s}{l^2-m_{\pi}^2+i\epsilon}\quad\frac{k^lk^t}{-\vec{k}^2}\quad\frac{1}{k^0-\frac{\vec{k}^2}{2M_N}+i\epsilon}\quad\frac{1}{k^0+\frac{\vec{k}^2}{2M_N}-i\epsilon}
\end{equation}
\begin{displaymath}
\times\frac{1}{k^0-l^0-\frac{\vec{k}^2}{2M_N}+i\epsilon}\quad\frac{1}{k^0-l^0+\frac{\vec{k}^2}{2M_N}-i\epsilon}
\ .
\end{displaymath}
This contribution, that is not the composition of one or two pion
potentials, nor does it vanish, is the new piece of the
nucleon-nucleon potential arising from considering non-relativistic (rather than static)
 nucleon propagators. Note that the key feature of this contribution is that one virtual energy ($l^0$) gets the scale $m_\pi$ from one of the loops and feeds it into the second loop so that the three-momentum of the latter ($\vec k$) gets in turn the scale $\sqrt{M_Nm_{\pi}}$. This large momentum scale makes the pion exchange interaction in that loop instantaneous. This is important in order to identify the relevant diagrams displayed in the Appendix, and also explains why at least two loops are necessary to see this kind of contributions.}

\end{itemize}

So, summarizing, after the analysis of poles of the propagator we have
found a purely three pion exchange potential (which is the result of
the static case), various iterations of one and two pion exchange
potentials, and a new contribution given by
\begin{eqnarray}
I^{new}&=&i\left(\frac{g_A}{F_{\pi}}\right)^6(7\vec{\tau}_1\cdot\vec{\tau}_2-6)\frac{\sigma_1^i}{2}\frac{\sigma_1^j}{2}\frac{\sigma_1^l}{2}\frac{\sigma_2^r}{2}\frac{\sigma_2^s}{2}\frac{\sigma_2^t}{2}\int\frac{d^dk}{(2\pi)^d}\int\frac{d^dl}{(2\pi)^d}\frac{k^ik^r}{-\vec{k}^2}\quad\frac{l^jl^s}{l^2-m_{\pi}^2+i\epsilon}\quad\frac{k^lk^t}{-\vec{k}^2} \label{eq} \nonumber\\
&
&\times\frac{1}{k^0-\frac{\vec{k}^2}{2M_N}+i\epsilon}\quad\frac{1}{k^0+\frac{\vec{k}^2}{2M_N}-i\epsilon}\quad\frac{1}{k^0-l^0-\frac{\vec{k}^2}{2M_N}+i\epsilon}\quad\frac{1}{k^0-l^0+\frac{\vec{k}^2}{2M_N}-i\epsilon}
\ . 
\end{eqnarray}

The tensorial part of our integral reduces to
\begin{displaymath}
\left(\frac{\sigma_1^i}{2}\frac{\sigma_1^j}{2}\frac{\sigma_1^i}{2}\frac{\sigma_2^r}{2}\frac{\sigma_2^j}{2}\frac{\sigma_2^r}{2}+\frac{\sigma_1^i}{2}\frac{\sigma_1^j}{2}\frac{\sigma_1^l}{2}\frac{\sigma_2^i}{2}\frac{\sigma_2^j}{2}\frac{\sigma_2^l}{2}+\frac{\sigma_1^i}{2}\frac{\sigma_1^j}{2}\frac{\sigma_1^l}{2}\frac{\sigma_2^l}{2}\frac{\sigma_2^j}{2}\frac{\sigma_2^i}{2}\right)\frac{\vec{k}^4\vec{l}^2}{(d-1)^2(d+1)}
\end{displaymath}
\begin{equation}
=\frac{1}{d-1}\frac{\vec{\sigma_1}\cdot\vec{\sigma_2}}{2^6}\vec{k}^4\vec{l}^2
\ .
\end{equation}
And so (\ref{eq}) becomes
\begin{eqnarray}
I^{new}&=&i\left(\frac{g_A}{F_{\pi}}\right)^6(7\vec{\tau}_1\cdot\vec{\tau}_2-6)\frac{1}{d-1}\frac{\vec{\sigma_1}\cdot\vec{\sigma_2}}{64}\int\frac{d^{d-1}\vec{k}}{(2\pi)^{d-1}}\int\frac{d^{d-1}\vec{l}}{(2\pi)^{d-1}}\vec{k}^4\vec{l^2} \nonumber \\
&
&\times\quad\frac{M_N^3}{\vec{k}^8}\quad\frac{1}{\sqrt{\vec{l}^2+m_{\pi}^2}}\quad\frac{1}{\vec{k}^2+M_N\sqrt{\vec{l}^2+m_{\pi}^2}}
\ .
\end{eqnarray}
Integrating over the solid angle,
\begin{equation}
\int d\Omega_d=\frac{2\pi^{d/2}}{\Gamma(d/2)} \ ,
\end{equation}
and defining
\begin{equation}
x=\frac{M_N\sqrt{\vec{l}^2+m_{\pi}^2}}{\vec{k}^2+M_N\sqrt{\vec{l}^2+m_{\pi}^2}} \quad \text{and} \quad y=\frac{m_{\pi}^2}{\vec{l}^2+m_{\pi}^2} \ ,
\end{equation}
we can transform the integrals into the product of two beta functions, and so we find that the new contribution is
\begin{eqnarray}
I^{new}&=&i\left(\frac{g_A}{F_{\pi}}\right)^6(7\vec{\tau}_1\cdot\vec{\tau}_2-6)\frac{1}{d-1}\frac{\vec{\sigma_1}\cdot\vec{\sigma_2}}{64}\frac{M_N^{\frac{d-1}{2}}m_{\pi}^{\frac{3d-7}{2}}}{2^{2d-2}\pi^{d-1}\left(\Gamma\left(\frac{d-1}{2}\right)\right)^2} \nonumber \\
& &
\times\frac{\Gamma\left(\frac{d+1}{2}\right)\Gamma\left(\frac{7-3d}{4}\right)}{\Gamma\left(\frac{9-d}{4}\right)}\Gamma\left(\frac{d-5}{2}\right)\Gamma\left(\frac{7-d}{2}\right)
\ .
\end{eqnarray}
Regularizing in $d=4-2\epsilon$ dimensions this is
\begin{eqnarray}
I^{new}&=&i\left(\frac{g_A}{F_{\pi}}\right)^6(7\vec{\tau}_1\cdot\vec{\tau}_2-6)\frac{1}{3}\frac{\vec{\sigma_1}\cdot\vec{\sigma_2}}{64}\frac{M_N^{\frac{3}{2}}m_{\pi}^{\frac{5}{2}}}{64\pi^3\left(\Gamma\left(\frac{3}{2}\right)\right)^2}\frac{\Gamma\left(\frac{5}{2}\right)\Gamma\left(\frac{-5}{4}\right)}{\Gamma\left(\frac{5}{4}\right)}\Gamma\left(\frac{-1}{2}\right)\Gamma\left(\frac{3}{2}\right) \nonumber \\
&=&-i\left(\frac{g_A}{F_{\pi}}\right)^6\frac{\Gamma\left(\frac{3}{4}\right)^2}{320\sqrt{2}\,\pi^{7/2}}M_N^{\frac{3}{2}}m_{\pi}^{\frac{5}{2}}(7\vec{\tau}_1\cdot\vec{\tau}_2-6)(\vec{\sigma_1}\cdot\vec{\sigma_2})
\ .
\end{eqnarray}

\section{Results}
\label{secdr}
By repeating the analysis of the previous section, one can show that only the diagrams displayed in the Appendix produce extra contributions when non-relativistic propagators are used instead of static ones. The contribution of some of the diagrams containing two local vertices (from (\ref{NN})) can be extracted from \cite{Mehen:1999hz}. In particular we have checked that the ratio of the pieces proportional to fractional powers of the masses arising from the diagrams Fig. 1 b) and Fig. 1 f) in that reference agrees with ratio of (A.13)  and (A.16) in the Appendix, as it should. There are further diagrams involving two-pion vertices with contributions in this region, but they are
suppressed by powers of $m_\pi/M_N$  with respect to the ones displayed in the Appendix.
This is due to the fact that two-pion vertices go with time derivatives ($\sim m_\pi$) rather than with space ones ($\sim \sqrt{m_\pi M_N}$). Adding up all remaining contributions, and using the identities
\bea
N^\dagger\tau^a \bfsigma N N^\dagger\tau^a\bfsigma N &=& -3N^\dagger N N^\dagger N \nn\\
N^\dagger\tau^a N N^\dagger\tau^a N &=& -2N^\dagger N N^\dagger N-N^\dagger\bfsigma N N^\dagger\bfsigma N
\eea
we obtain the following contribution to the (momentum space) potential
\begin{equation}
V=
\frac{3\,\Gamma\left(\frac{3}{4}\right)^2}{10\sqrt{2}\,\pi^{7/2}}M_N^{\frac{3}{2}}m_{\pi}^{\frac{5}{2}}
\left\{{1\over 2} \left(\frac{g_A}{F_{\pi}}\right)^6
+\left(\frac{g_A}{F_{\pi}}\right)^2
\left[C_S^2
-C_SC_T\left(2(\vec{\sigma}_1\cdot\vec{\sigma}_2)+4\right)
+C_T^2\left( 2(\vec{\sigma}_1\cdot\vec{\sigma}_2) + 23 \right)
\right]
\right\} 
\end{equation}
This amounts to the following redefinition of $C_S$ and $C_T$
\begin{eqnarray}
\label{C_Sredef}
C_S\longrightarrow  \tilde{C_S}  &=&  C_S+
\frac{3\,\Gamma\left(\frac{3}{4}\right)^2}{10\sqrt{2}\,\pi^{7/2}}M_N^{\frac{3}{2}}m_{\pi}^{\frac{5}{2}}
\left\{{1\over 2}\left(\frac{g_A}{F_{\pi}}\right)^6+\left(\frac{g_A}{F_{\pi}}\right)^2\left(C_S^2-4C_SC_T+23C_T^2\right)
\right\} \nonumber\\
 &\equiv& C_S+\Delta C_S \\
\nonumber \\
\label{C_Tredef}
C_T\longrightarrow\tilde{C_T}&=&C_T+
\frac{3\,\Gamma\left(\frac{3}{4}\right)^2}{10\sqrt{2}\,\pi^{7/2}}M_N^{\frac{3}{2}}m_{\pi}^{\frac{5}{2}}
\left\{
2\left(\frac{g_A}{F_{\pi}}\right)^2\left(-C_SC_T+ C_T^2\right)
\right\} \nonumber \\
&\equiv&C_T+\Delta C_T
\end{eqnarray}

Hence the previous extractions from data of these coefficients must be corrected according to the formula above\footnote{There are further contributions from the iteration of the potentials and from higher order potentials themselves which eventually amount to redefinitions of $C_S$ and $C_T$. These contributions, however, have already been taken into account (see, for instance, \cite{Epelbaum:2004fk}) in the fits to data. We restrict our discussion to the new type of contributions found in this work.}. For instance, we may use the extraction found in ref. \cite{Epelbaum:2004fk}. There we may find different results for $C_S$ and $C_T$ in the $np$ channel coming from a fit done for different values of the cut-offs $\Lambda$ and $\tilde{\Lambda}$, which enter the Lippmann-Schwinger equation and the spectral-function representation of the two-pion exchange potential, respectively. The approximate results are shown in Table I.
\begin{table}[ht]
\begin{tabular}{| c | c | c | c | c|}
\hline
LEC & $ \{450,500\} $ & $ \{600,600\} $ & $ \{450,700\} $ & $ \{600,700\} $ \\
\hline
 $C_S$ & $ -10.7 $&$8.9 $ & $-12.1 $& $3.4 $\\
\hline
$C_T$ & $-1.2 $& $ 5.3 $& $-0.6 $& $2.5 $\\
\hline
\end{tabular}
\caption{The approximate values of the S-wave LECs $C_S$ and $C_T$ in the $np$ channel at $N^3LO$ for the different cut-off combinations $ \{\Lambda[\text{MeV}],\tilde{\Lambda}[\text{MeV}]\} $, from the fit in \cite{Epelbaum:2004fk}. The values of the constants are in $10^{-5}\ \text{MeV}^{-2}$.}
\end{table}

If we use these values of $C_S$ and $C_T$ to find an estimate of the size of the corrections $\Delta C_S$ and $\Delta C_T$, we find the values displayed in Table II.
\begin{table}[hh]
\begin{tabular}{| c | c | c | c | c|}
\hline
$\Delta$ & $ \{450,500\} $ & $ \{600,600\} $ & $ \{450,700\} $ & $ \{600,700\} $ \\
\hline
$\Delta C_S$ & $ 21.2 $ & $ 54.1 $ & $ 23.3 $ & $ 23.5 $\\
\hline
$\Delta C_T$ & $ -1.6 $ & $ -2.8 $ & $ -1.0 $ & $ -0.3 $\\
\hline
\end{tabular}
\caption{The values of the corrections $\Delta C_S$ and $\Delta C_T$ calculated using $C_S$ and $C_T$ from the former table as input. We use $g_A=1.29$, $F_{\pi}=92.4$ MeV, $M_N=939$ MeV, and $m_{\pi}=139$ MeV. The values of the constants are in $10^{-5}\ \text{MeV}^{-2}$.}
\end{table}
Clearly these are no small corrections. In fact, if using the data from \cite{Epelbaum:2004fk} we solve equations (\ref{C_Sredef}) and (\ref{C_Tredef}) exactly, we find complex values for $C_S$ and $C_T$ in all cases. This would indicate the need to redo the fits in \cite{Epelbaum:2004fk} taking into account these new contributions.

\section{Discussion}
\label{secd}

At first sight the new contributions to the potential we have found may look irrelevant since they amount to redefinitions of local counterterms. This is probably so as far as the description of scattering data is concerned. However, they may be of practical importance at least in the following two issues: (i)  they shift the values of $C_S$ and $C_T$ extracted from data, which is important in order to check the consistency of a given counting scheme, and (ii) they indicate that there will be contributions going like $m_q^{5/4}$, $m_q$ being the light quark masses, which should be taken into account if one aims at a precision calculation of the nucleon-nucleon scattering lengths from the lattice QCD using chiral extrapolations \cite{Kaplan:2006sv}.

Notice that the scale $\sqrt{m_\pi M_N}$ coincides (up to corrections $m_\pi/ M_N$) with the minimum momentum of the nucleons to produce a pion at rest. This coincidence suggests the possibility of an alternative derivation of our results, by computing the three-body spectral function and then making use of a dispersion relation. 

In higher loop calculations, subdiagrams with nucleon energies scaling as $m_\pi$ and nucleon three-momentum as $\sqrt{m_\pi M_N}$ will appear. It is not difficult to convince oneself that such subdiagrams are produced by adding an extra one pion exchange or a contact term to the two-loop diagrams we have calculated. These additions amount to a suppression of a factor $M_N^{3/2}m_\pi^{1/2}/\Lambda_\chi^2$, which, parametrically, is equivalent to $m_\pi^{1/2}/M_N^{1/2}$ if $\Lambda_\chi \sim M_N$. However, in practice $\Lambda_\chi^2$ takes a value $\sim 16\pi F_\pi^2/g_A^2$ in these contributions, which makes the ratio $M_N^{3/2}m_\pi^{1/2}/\Lambda_\chi^2\sim 1.2$. Hence, one may consider a modified power counting in which $M_N^{3/2}m_\pi^{1/2}/\Lambda_\chi^2\sim 1$, so that all these higher loop calculations should be consistently summed up. Whether this is feasible or not and whether this might provide an explanation to the unnatural size of the scattering lengths will be left for future work.

Analogous contributions, proportional to fractional powers of the nucleon mass, are also expected in models with (arbitrary) boson exchanges at two loops (and beyond), as recently considered in \cite{Kaiser:2006yh}. In this reference non-relativistic propagators are used to calculate the contributions to the amplitude coming from the iterations of one-boson-exchange and two-boson-exchange potentials, but not in the calculation of the three-boson-exchange potential itself, for which the static approximation is used. We believe that, if non-relativistic rather than static propagators were used for the latter, contributions with the above mentioned fractional powers of the nucleon mass would also be obtained.


\section{Conclusions}
\label{secc}
Using the method of the threshold expansions, we have found contributions to the nucleon-nucleon potential which are missed if static rather than non-relativistic nucleon propagators are used in the calculation and have calculated the leading contributions of them, which appear at two loops. They produce large contact terms with a peculiar non-analytic dependence on the light quark masses. 

\bigskip

{\bf Acknowledgements.}

\bigskip

JS thanks Harald Griesshammer and Daniel Phillips for important discussions. He is also thankful to the organizers of ECT* activity {\it Effective Theories in Nuclear Physics and Lattice QCD}, where these discussions took place and part of this work was carried out. JM is supported by a \textit{Distinci\'o} from the \textit{Generalitat de Catalunya}. We acknowledge financial support from MEC (Spain) grant CYT FPA
2004-04582-C02-01, the CIRIT (Catalonia) grant 2005SGR00564,  and the RTNs
Euridice HPRN-CT2002-00311 and Flavianet MRTN-CT-2006-035482 (EU).


\appendix

\section{Results for individual diagrams}
\label{app}

\subsection{Diagrams with no contact terms}

\begin{eqnarray}
\parbox{30mm}{
\begin{fmfgraph*}(150,60)
\fmfleft{i1,i2} \fmfright{o1,o2} \fmf{heavy}{i2,v1}
\fmf{heavy}{v1,v2} \fmf{heavy}{v2,v3} \fmf{heavy}{v3,o2}
\fmf{heavy}{i1,u1} \fmf{heavy}{u1,u2} \fmf{heavy}{u2,u3}
\fmf{heavy}{u3,o1} \fmffreeze \fmf{fermion}{u1,v1}
\fmf{fermion}{v2,u2} \fmf{fermion}{u3,v3}
\end{fmfgraph*}}\quad\quad\quad\quad\quad
&=&-i\left(\frac{g_A}{F_{\pi}}\right)^6\frac{\Gamma\left(\frac{3}{4}\right)^2}{320\sqrt{2}\,\pi^{7/2}}M_N^{\frac{3}{2}}m_{\pi}^{\frac{5}{2}}(7\vec{\tau}_1\cdot\vec{\tau}_2-6)(\vec{\sigma}_1\cdot\vec{\sigma}_2) \nonumber \\
 \\
\nonumber \\
\nonumber \\
\parbox{30mm}
{\begin{fmfgraph*}(150,60) \fmfleft{i1,i2} \fmfright{o1,o2}
\fmf{double}{i1,v1,v2,v3,o1} \fmf{double}{i2,u1,u2,u3,o2} \fmffreeze
\fmf{heavy}{i2,u1} \fmf{heavy}{i1,v1} \fmf{heavy}{u3,o2}
\fmf{heavy}{v3,o1} \fmf{heavy}{v1,v2} \fmf{heavy}{v2,v3}
\fmf{heavy}{u1,u2} \fmf{heavy}{u2,u3} \fmf{vanilla}{v2,x,u1}
\fmf{fermion}{x,u1} \fmf{vanilla}{u2,y,v1} \fmf{fermion}{y,v1}
\fmf{fermion}{v3,u3}
\end{fmfgraph*}}\quad\quad\quad\quad\quad
&=&i\left(\frac{g_A}{F_{\pi}}\right)^6\frac{\Gamma\left(\frac{3}{4}\right)^2}{960\sqrt{2}\,\pi^{7/2}}M_N^{\frac{3}{2}}m_{\pi}^{\frac{5}{2}}(\vec{\tau}_1\cdot\vec{\tau}_2-6)(\vec{\sigma}_1\cdot\vec{\sigma}_2) \nonumber \\
 \\
\nonumber \\
\nonumber \\
\parbox{30mm}
{\begin{fmfgraph*}(150,60) \fmfleft{i1,i2} \fmfright{o1,o2}
\fmf{double}{i1,v1,v2,v3,o1} \fmf{double}{i2,u1,u2,u3,o2} \fmffreeze
\fmf{heavy}{i2,u1} \fmf{heavy}{u1,u2} \fmf{heavy}{u2,u3}
\fmf{heavy}{u3,o2} \fmf{heavy}{i1,v1} \fmf{heavy}{v1,v2}
\fmf{heavy}{v2,v3} \fmf{heavy}{v3,o1} \fmf{fermion}{v1,u3}
\fmf{fermion}{u2,v3} \fmf{fermion}{v2,u1}
\end{fmfgraph*}}\quad\quad\quad\quad\quad
&=&i\left(\frac{g_A}{F_{\pi}}\right)^6\frac{\Gamma\left(\frac{3}{4}\right)^2}{640\sqrt{2}\,\pi^{7/2}}M_N^{\frac{3}{2}}m_{\pi}^{\frac{5}{2}}(\vec{\tau}_1\cdot\vec{\tau}_2+6)(\vec{\sigma}_1\cdot\vec{\sigma}_2) \nonumber \\
 \\
\nonumber \\
\nonumber \\
\parbox{30mm}
{\begin{fmfgraph*}(150,60) \fmfleft{i1,i2} \fmfright{o1,o2}
\fmf{double}{i1,v1,v2,v3,v4,o1} \fmf{double}{i2,u1,u2,u3,u4,o2}
\fmffreeze \fmf{heavy}{i1,v1} \fmf{heavy}{v2,v3} \fmf{heavy}{v4,o1}
\fmf{heavy}{i2,u1} \fmf{heavy}{u1,u2} \fmf{heavy}{u2,u3}
\fmf{heavy}{u3,u4} \fmf{heavy}{u4,o2} \fmf{fermion}{u2,v2}
\fmf{fermion}{v3,u3} \fmf{fermion,left=0.7,tension=0.5}{u1,u4}
\end{fmfgraph*}}\quad\quad\quad\quad\quad
&=&-i\left(\frac{g_A}{F_{\pi}}\right)^6\frac{3\,\Gamma\left(\frac{3}{4}\right)^2}{640\sqrt{2}\,\pi^{7/2}}M_N^{\frac{3}{2}}m_{\pi}^{\frac{5}{2}}\left(2\vec{\tau}_1\cdot\vec{\tau}_2+9\right)
 \\
\nonumber \\
\nonumber \\
\nonumber \\
\parbox{30mm}{
\begin{fmfgraph*}(150,60)
\fmfleft{i1,i2} \fmfright{o1,o2}
\fmf{double}{i1,v1,vx,v2,v3,vy,v4,o1}
\fmf{double}{i2,u1,ux,u2,u3,uy,u4,o2} \fmffreeze \fmf{heavy}{i1,v1}
\fmf{heavy}{v4,o1} \fmf{heavy}{v2,v3} \fmf{heavy}{i2,u1}
\fmf{heavy}{u2,u3} \fmf{heavy}{u4,o2} \fmf{fermion}{u1,v1}
\fmf{fermion}{v4,u4} \fmf{fermion,left}{ux,uy}
\end{fmfgraph*}}\quad\quad\quad\quad\quad
&=&i\left(\frac{g_A}{F_{\pi}}\right)^6\frac{3\,\Gamma\left(\frac{3}{4}\right)^2}{640\sqrt{2}\,\pi^{7/2}}M_N^{\frac{3}{2}}m_{\pi}^{\frac{5}{2}}(6\vec{\tau}_1\cdot\vec{\tau}_2-9)
 \\
\nonumber \\
\nonumber \\
\nonumber \\
\parbox{30mm}
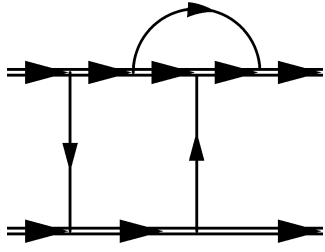
{\begin{fmfgraph*}(150,60) \fmfleft{i1,i2} \fmfright{o1,o2}
\fmf{phantom}{i1,v1,v2,v3,v4,o1} \fmf{phantom}{i2,u1,u2,u3,u4,o2}
\fmffreeze \fmf{heavy}{i1,v1} \fmf{heavy}{v1,v3} \fmf{double}{v3,v4}
\fmf{heavy}{v4,o1} \fmf{heavy}{i2,u1} \fmf{heavy}{u1,u2}
\fmf{heavy}{u2,u3} \fmf{heavy}{u3,u4} \fmf{heavy}{u4,o2}
\fmf{fermion}{u1,v1} \fmf{fermion}{v3,u3} \fmf{fermion,left}{u2,u4}
\end{fmfgraph*}}\quad\quad\quad\quad\quad
&=&-i\left(\frac{g_A}{F_{\pi}}\right)^6\frac{\Gamma\left(\frac{3}{4}\right)^2}{640\sqrt{2}\,\pi^{7/2}}M_N^{\frac{3}{2}}m_{\pi}^{\frac{5}{2}}(2\vec{\tau}_1\cdot\vec{\tau}_2-3)
\end{eqnarray}

\subsection{Diagrams with one contact term}

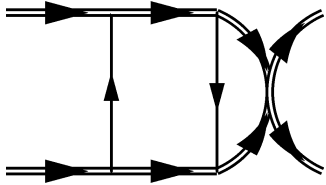
\begin{eqnarray}
\parbox{30mm}{
\begin{fmfgraph*}(150,60)
\fmfleft{i1,i2} \fmfright{o1,o2} \fmf{heavy}{i1,u1,u2}
\fmf{heavy}{i2,v1,v2} \fmf{phantom}{u2,o1} \fmf{phantom}{v2,o2}
\fmffreeze \fmf{heavy,right=0.3}{u2,x} \fmf{heavy,left=0.3}{x,o2}
\fmf{heavy,left=0.3}{v2,x} \fmf{heavy,right=0.3}{x,o1}
\fmf{heavy}{u1,u2} \fmf{heavy}{v1,v2} \fmf{fermion}{u1,v1}
\fmf{fermion}{v2,u2}
\end{fmfgraph*}}\quad\quad\quad\quad\quad
&=&i\left(\frac{g_A}{F_{\pi}}\right)^4\frac{\Gamma\left(\frac{3}{4}\right)^2}{240\sqrt{2}\,\pi^{7/2}}M_N^{\frac{3}{2}}m_{\pi}^{\frac{5}{2}}(2\vec{\tau}_1\cdot\vec{\tau}_2-3) \nonumber \\
& & \times \left(C_S(2\vec{\sigma}_1\cdot\vec{\sigma}_2-3)-C_T(7\vec{\sigma}_1\cdot\vec{\sigma}_2-6)\right)\\
\nonumber \\
\nonumber \\
\parbox{30mm}
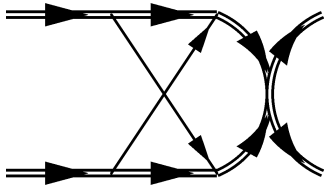
{
\begin{fmfgraph*}(150,60)
\fmfleft{i1,i2} \fmfright{o1,o2} \fmf{heavy}{i1,u1,u2}
\fmf{heavy}{i2,v1,v2} \fmf{phantom}{u2,o1} \fmf{phantom}{v2,o2}
\fmffreeze \fmf{heavy,right=0.3}{u2,x} \fmf{heavy,left=0.3}{x,o2}
\fmf{heavy,left=0.3}{v2,x} \fmf{heavy,right=0.3}{x,o1}
\fmf{heavy}{u1,u2} \fmf{heavy}{v1,v2} \fmf{vanilla}{u1,y,v2}
\fmf{fermion}{y,v2} \fmf{vanilla}{v1,z,u2} \fmf{fermion}{z,u2}
\end{fmfgraph*}}\quad\quad\quad\quad\quad
&=&-i\left(\frac{g_A}{F_{\pi}}\right)^4\frac{\Gamma\left(\frac{3}{4}\right)^2}{240\sqrt{2}\,\pi^{7/2}}M_N^{\frac{3}{2}}m_{\pi}^{\frac{5}{2}}(2\vec{\tau}_1\cdot\vec{\tau}_2+3) \nonumber \\
& &\times(C_S(2\vec{\sigma}_1\cdot\vec{\sigma}_2+3)-C_T(\vec{\sigma}_1\cdot\vec{\sigma}_2-6)) \\
\nonumber \\
\nonumber \\
\parbox{30mm}
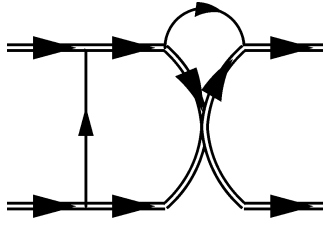
{
\begin{fmfgraph*}(150,60)
\fmfleft{i1,i2} \fmfright{o1,o2} \fmf{phantom}{i1,u1,u2,u3,o1}
\fmf{phantom}{i2,v1,v2,v3,o2} \fmffreeze \fmf{heavy}{i1,u1,u2}
\fmf{heavy}{i2,v1,v2} \fmf{heavy,left=0.2}{v2,x}
\fmf{heavy,left=0.2}{x,v3} \fmf{double,right=0.2}{u2,x}
\fmf{double,right=0.2}{x,u3} \fmf{heavy}{v3,o2} \fmf{heavy}{u3,o1}
\fmf{fermion}{u1,v1} \fmf{fermion,left}{v2,v3}
\end{fmfgraph*}}\quad\quad\quad\quad\quad
&=&-i\left(\frac{g_A}{F_{\pi}}\right)^4\frac{\Gamma\left(\frac{3}{4}\right)^2}{480\sqrt{2}\,\pi^{7/2}}M_N^{\frac{3}{2}}m_{\pi}^{\frac{5}{2}} \nonumber \\
& &\times\left[(3\vec{\tau}_1\cdot\vec{\tau}_2)\left(3C_S(\vec{\sigma}_1\cdot\vec{\sigma}_2)+C_T(2\vec{\sigma}_1\cdot\vec{\sigma}_2-3)\right) \right. \nonumber\\ 
& &\left.+(2\vec{\tau}_1\cdot\vec{\tau}_2-3)\left(C_S(2\vec{\sigma}_1\cdot\vec{\sigma}_2-3)+C_T(\vec{\sigma}_1\cdot\vec{\sigma}_2-6)\right)\right] \\
\nonumber \\
\nonumber \\
\parbox{30mm}
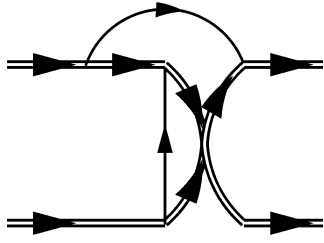
{
\begin{fmfgraph*}(150,60)
\fmfleft{i1,i2} \fmfright{o1,o2} \fmf{phantom}{i1,u1,u2,u3,o1}
\fmf{phantom}{i2,v1,v2,v3,o2} \fmffreeze \fmf{double}{i1,u1,u2}
\fmf{heavy}{i1,u1} \fmf{heavy}{i2,v1,v2} \fmf{heavy,left=0.2}{v2,x}
\fmf{heavy,left=0.2}{x,v3} \fmf{heavy,right=0.2}{u2,x}
\fmf{double,right=0.2}{x,u3} \fmf{heavy}{v3,o2} \fmf{heavy}{u3,o1}
\fmf{fermion}{u2,v2} \fmf{fermion,left=0.7}{v1,v3}
\end{fmfgraph*}}\quad\quad\quad\quad\quad
&=&i\left(\frac{g_A}{F_{\pi}}\right)^4\frac{\Gamma\left(\frac{3}{4}\right)^2}{480\sqrt{2}\,\pi^{7/2}}M_N^{\frac{3}{2}}m_{\pi}^{\frac{5}{2}}\nonumber \\
& &\times\left[(\vec{\tau}_1\cdot\vec{\tau}_2)\left(C_S(\vec{\sigma}_1\cdot\vec{\sigma}_2)-C_T(2\vec{\sigma}_1\cdot\vec{\sigma}_2+9)\right) \right. \nonumber \\
& & \left. +(2\vec{\tau}_1\cdot\vec{\tau}_2+3)\left(C_S(2\vec{\sigma}_1\cdot\vec{\sigma}_2+3)-C_T(\vec{\sigma}_1\cdot\vec{\sigma}_2+6)\right)\right] \\
\nonumber \\
\nonumber \\
\parbox{30mm}
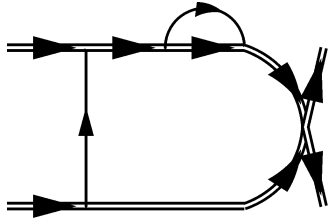
{
\begin{fmfgraph*}(150,60)
\fmfleft{i1,i2} \fmfright{o1,o2} \fmf{phantom}{i1,u1,u2,u3,o1}
\fmf{phantom}{i2,v1,v2,v3,o2} \fmffreeze \fmf{heavy}{i2,v1,v2,v3}
\fmf{double}{i1,u1,u2,u3} \fmf{heavy}{i1,u1} \fmf{fermion}{u1,v1}
\fmf{fermion,left}{v2,v3} \fmf{heavy,left=0.3,tension=0.3}{v3,x}
\fmf{heavy}{x,o1} \fmf{heavy,right=0.3,tension=0.3}{u3,x}
\fmf{heavy}{x,o2}
\end{fmfgraph*}}\quad\quad\quad\quad\quad
&=&i\left(\frac{g_A}{F_{\pi}}\right)^4\frac{3\,\Gamma\left(\frac{3}{4}\right)^2}{160\sqrt{2}\,\pi^{7/2}}M_N^{\frac{3}{2}}m_{\pi}^{\frac{5}{2}}(\vec{\tau}_1\cdot\vec{\tau}_2)(C_S(\vec{\sigma}_1\cdot\vec{\sigma}_2)-C_T(2\vec{\sigma}_1\cdot\vec{\sigma}_2-3)) \nonumber \\
\\
\nonumber \\
\nonumber \\
\parbox{30mm}
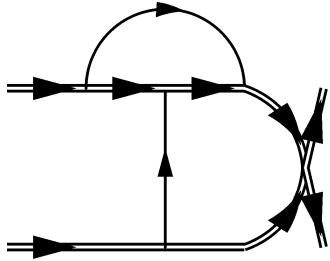
{
\begin{fmfgraph*}(150,60)
\fmfleft{i1,i2} \fmfright{o1,o2} \fmf{phantom}{i1,u1,u2,u3,o1}
\fmf{phantom}{i2,v1,v2,v3,o2} \fmffreeze \fmf{heavy}{i2,v1,v2,v3}
\fmf{double}{i1,u1,u2,u3} \fmf{heavy}{i1,u1} \fmf{fermion}{u2,v2}
\fmf{fermion,left}{v1,v3} \fmf{heavy,left=0.3,tension=0.3}{v3,x}
\fmf{heavy}{x,o1} \fmf{heavy,right=0.3,tension=0.3}{u3,x}
\fmf{heavy}{x,o2}
\end{fmfgraph*}}\quad\quad\quad\quad\quad
&=&-i\left(\frac{g_A}{F_{\pi}}\right)^4\frac{\Gamma\left(\frac{3}{4}\right)^2}{480\sqrt{2}\,\pi^{7/2}}M_N^{\frac{3}{2}}m_{\pi}^{\frac{5}{2}}(\vec{\tau}_1\cdot\vec{\tau}_2)\left(C_S\vec{\sigma}_1\cdot\vec{\sigma}_2-C_T(2\vec{\sigma}_1\cdot\vec{\sigma}_2-3)\right) \nonumber \\
\nonumber \\
\end{eqnarray}

\subsection{Diagrams with two contact terms}

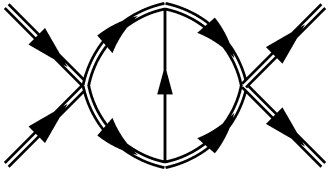
\begin{eqnarray}
\parbox{30mm}{
\begin{fmfgraph*}(150,60)
\fmfleft{i1,i2} \fmfright{o1,o2} \fmf{phantom}{i1,u,o1}
\fmf{phantom}{i2,v,o2} \fmffreeze \fmf{heavy}{i1,x}
\fmf{heavy}{i2,x} \fmf{heavy,right=0.3}{x,u}
\fmf{heavy,left=0.3}{x,v} \fmf{fermion}{u,v}
\fmf{heavy,right=0.3}{u,y} \fmf{heavy,left=0.3}{v,y}
\fmf{heavy}{y,o1} \fmf{heavy}{y,o2}
\end{fmfgraph*}}\quad\quad\quad\quad\quad
&=&-i\left(\frac{g_A}{F_{\pi}}\right)^2\frac{\Gamma\left(\frac{3}{4}\right)^2}{20\sqrt{2}\,\pi^{7/2}}M_N^{\frac{3}{2}}m_{\pi}^{\frac{5}{2}}(\vec{\tau}_1\cdot\vec{\tau}_2)\nonumber \\
& &\times\left(C_S^2\vec{\sigma}_1\cdot\vec{\sigma}_2-2C_SC_T(2\vec{\sigma}_1\cdot\vec{\sigma}_2-3)+C_T^2(7\vec{\sigma}_1\cdot\vec{\sigma}_2-6)\right) \\
\nonumber \\
\nonumber \\
\nonumber \\
\parbox{30mm}
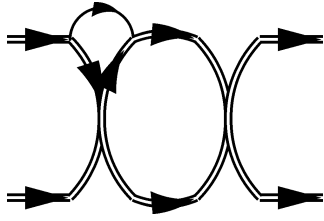
{
\begin{fmfgraph*}(150,60)
\fmfleft{i1,i2} \fmfright{o1,o2} \fmf{phantom}{i1,u1,u2,u3,u4,o1}
\fmf{phantom}{i2,v1,v2,v3,v4,o2} \fmffreeze \fmf{heavy}{i1,u1}
\fmf{heavy}{i2,v1} \fmf{double,right=0.2}{u1,x}
\fmf{heavy,left=0.2}{v1,x} \fmf{double,right=0.2}{x,u2}
\fmf{heavy,left=0.2}{x,v2} \fmf{heavy,right=0.2}{u2,u3}
\fmf{heavy,left=0.2}{v2,v3} \fmf{double,right=0.2}{u3,y}
\fmf{double,left=0.2}{v3,y} \fmf{double,right=0.2}{y,u4}
\fmf{double,left=0.2}{y,v4} \fmf{heavy}{u4,o1} \fmf{heavy}{v4,o2}
\fmffreeze \fmf{fermion,left}{v1,v2}
\end{fmfgraph*}}\quad\quad\quad\quad\quad
&=&i\left(\frac{g_A}{F_{\pi}}\right)^2\frac{\Gamma\left(\frac{3}{4}\right)^2}{40\sqrt{2}\,\pi^{7/2}}M_N^{\frac{3}{2}}m_{\pi}^{\frac{5}{2}}\nonumber \\
& &\times\left[3(3C_S^2+2C_SC_T(\vec{\sigma}_1\cdot\vec{\sigma}_2)+C_T^2(2\vec{\sigma}_1\cdot\vec{\sigma}_2-3)) \right. \nonumber\\
&
&\left.+(\vec{\tau}_1\cdot\vec{\tau}_2)(C_S^2(\vec{\sigma}_1\cdot\vec{\sigma_2})+6C_SC_T-C_T^2(\vec{\sigma}_1\cdot\vec{\sigma}_2-6))\right] \\
\nonumber \\
\nonumber \\
\parbox{30mm}
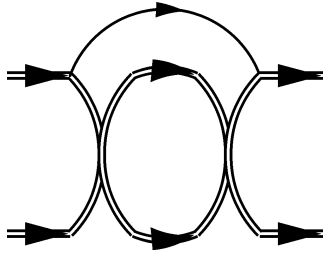
{
\begin{fmfgraph*}(150,60)
\fmfleft{i1,i2} \fmfright{o1,o2} \fmf{phantom}{i1,u1,u2,u3,u4,o1}
\fmf{phantom}{i2,v1,v2,v3,v4,o2} \fmffreeze \fmf{heavy}{i1,u1}
\fmf{heavy}{i2,v1} \fmf{double,right=0.2}{u1,x}
\fmf{double,left=0.2}{v1,x} \fmf{double,right=0.2}{x,u2}
\fmf{double,left=0.2}{x,v2} \fmf{heavy,right=0.2}{u2,u3}
\fmf{heavy,left=0.2}{v2,v3} \fmf{double,right=0.2}{u3,y}
\fmf{double,left=0.2}{v3,y} \fmf{double,right=0.2}{y,u4}
\fmf{double,left=0.2}{y,v4} \fmf{heavy}{u4,o1} \fmf{heavy}{v4,o2}
\fmffreeze \fmf{fermion,left=0.7}{v1,v4}
\end{fmfgraph*}}\quad\quad\quad\quad\quad
&=&-i\left(\frac{g_A}{F_{\pi}}\right)^2\frac{\Gamma\left(\frac{3}{4}\right)^2}{20\sqrt{2}\,\pi^{7/2}}M_N^{\frac{3}{2}}m_{\pi}^{\frac{5}{2}}\nonumber \\
& &\times\left[3(3C_S^2-2C_SC_T(\vec{\sigma}_1\cdot\vec{\sigma}_2)+C_T^2(2\vec{\sigma}_1\cdot\vec{\sigma}_2+9))\right. \nonumber \\
& &\left.+(\vec{\tau}_1\cdot\vec{\tau}_2)\left(C_S^2(\vec{\sigma}_1\cdot\vec{\sigma}_2)+2C_SC_T(2\vec{\sigma}_1\cdot\vec{\sigma}_2+3)-C_T^2(\vec{\sigma}_1\cdot\vec{\sigma}_2+6)\right)\right] \nonumber \\
\\
\nonumber \\
\nonumber \\
\parbox{30mm}
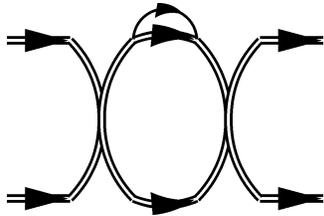
{
\begin{fmfgraph*}(150,60)
\fmfleft{i1,i2} \fmfright{o1,o2} \fmf{phantom}{i1,u1,u2,u3,u4,o1}
\fmf{phantom}{i2,v1,v2,v3,v4,o2} \fmffreeze \fmf{heavy}{i1,u1}
\fmf{heavy}{i2,v1} \fmf{double,right=0.2}{u1,x}
\fmf{double,left=0.2}{v1,x} \fmf{double,right=0.2}{x,u2}
\fmf{double,left=0.2}{x,v2} \fmf{heavy,right=0.2}{u2,u3}
\fmf{heavy,left=0.2}{v2,v3} \fmf{double,right=0.2}{u3,y}
\fmf{double,left=0.2}{v3,y} \fmf{double,right=0.2}{y,u4}
\fmf{double,left=0.2}{y,v4} \fmf{heavy}{u4,o1} \fmf{heavy}{v4,o2}
\fmffreeze \fmf{fermion,left}{v2,v3}
\end{fmfgraph*}}\quad\quad\quad\quad\quad
&=&-i\left(\frac{g_A}{F_{\pi}}\right)^2\frac{9\,\Gamma\left(\frac{3}{4}\right)^2}{40\sqrt{2}\,\pi^{7/2}}M_N^{\frac{3}{2}}m_{\pi}^{\frac{5}{2}}(C_S^2+2C_SC_T(\vec{\sigma}_1\cdot\vec{\sigma}_2)-C_T^2(2\vec{\sigma}_1\vec{\sigma}_2-3)) \nonumber \\
\end{eqnarray}


\end{fmffile}

\end{document}